\documentclass[12pt,]{iopart}
\usepackage{iopams}  

\usepackage{bm,mathrsfs,color}
\usepackage{graphicx}
\usepackage{braket}

\newcommand{\bt}{\bm{t}}
\newcommand{\bu}{\bm{u}}

\newcommand{\kp}{\bm{k}\cdot \bm{p}}

\newcommand{\wc}{\omega_c}

\newcommand{\eqref}[1]{(\ref{#1})}

\begin{document}

\title[Crystalline spin-orbit interaction and the Zeeman splitting in Pb$_{1-x}$Sn$_x$Te]{
%\underline{\small{\tt ver. 1.1 (\today)}}
%\vspace{5mm}\\
Crystalline spin-orbit interaction and the Zeeman splitting in Pb$_{1-x}$Sn$_x$Te}

\author{Hiroshi Hayasaka and Yuki Fuseya}

\address{Department of Engineering Science, University of Electro-Communications, Chofu, Tokyo 182-8585, Japan}
\ead{fuseya@uec.ac.jp}
\vspace{10pt}
%\begin{indented}
%\item[]February 2014
%\end{indented}

\begin{abstract}
The ratio of the Zeeman splitting to the cyclotron energy ($M=\Delta E_Z / \hbar \wc$), which characterizes the relative strength of the spin-orbit interaction in crystals, is examined for the narrow gap IV-VI semiconductors PbTe, SnTe, and their alloy Pb$_{1-x}$Sn$_x$Te on the basis of the multiband $\kp$ theory. The inverse mass $\alpha$, the g-factor $g$, and $M$ are calculated numerically by employing the relativistic empirical tight-binding band calculation. On the other hand, a simple but exact formula of $M$ is obtained for the six-band model based on the group theoretical analysis. It is shown that $M<1$ for PbTe and $M>1$ for SnTe, which are interpreted in terms of the relevance of the interband couplings due to the crystalline spin-orbit interaction. It is clarified both analytically and numerically that $M=1$ just at the band inversion point, where the transition from trivial to nontrivial topological crystalline insulator occurs. By using this property, one can detect the transition point only with the bulk measurements. It is also proposed that $M$ is useful to evaluate quantitatively a degree of the Dirac electrons in solids.
\end{abstract}

%\maketitle

%\section{Introduction}
	The spin-orbit interaction (SOI) affects the eigenstate of electrons in solids in a variety of ways. It strongly depends on the crystal structure and the momentum of the carrier. One of the most fundamental such effects is the modification of the band structure. For example, in semiconductors of the diamond and zincblende structures, the band modification can be characterized by the spin-orbit splitting energy. But this is not the whole information of the crystalline SOI. Another important information can be obtained under a magnetic field, where we cannot attain to only with the band calculations. The one-body Hamiltonian under the magnetic field can be separated into two part in general: the symmetric and the antisymmetric part with respect to the commutation of the kinematical momentum operator\cite{Luttinger1956,Roth1959,Cohen1960,Yafet1963,Fuseya2015}. The eigenenergy of the  symmetric part is given in terms of the cyclotron energy as $\hbar \wc (n+1/2)$ with an anisotropic cyclotron mass. The eigenenergy of the antisymmetric part is given by the Zeeman energy with an anisotropic g-factor. This Zeeman energy does not originate from the bare electron spins, but originates from the orbital motion of electrons. The antisymmetric part is relevant only in the case with the sizable crystalline SOI. %except for the small contribution from the bare Zeeman energy. 
	Therefore, the effect of the crystalline SOI is clearly reflected by the antisymmetric part, whose relative strength is characterized by the ratio of the Zeeman splitting to the cyclotron energy, $M=\Delta E_Z / \hbar \wc$. What is important here is that the ratio $M$ is an intrinsic value, which is not affected by impurities or vacancies, and it can be accurately determined by the quantum oscillation measurements.

	So far, the ratio $M$ has been determined repeatedly by experiments in strongly spin-orbit coupled materials, such as Bi\cite{Smith1964,Edelman1976,Bompadre2001,Behnia2007_PRL,ZZhu2011} and Bi$_2$Se$_3$\cite{Kohler1975,Fauque2013,Orlita2015}. On the other hand, however, the clear theoretical understandings of $M$ has been pushed aside except for the two-band Dirac electron systems\cite{Cohen1960,Wolff1964,Fuseya2015}. %Especially, how the crystalline SOI modifies the Zeeman splitting has not been substantiated except for the Dirac electron systems\cite{Cohen1960,Wolff1964,Fuseya2015}. 
	Actually, the anisotropic and large $M$ in Bi has been a puzzle for more than half a century\cite{Smith1964,Edelman1976,Bompadre2001,Behnia2007_PRL,ZZhu2011}. Very recently, this long-standing puzzle was eventually solved based on the multiband $\kp$ theory\cite{Fuseya2015b}. It was newly clarified that the ratio $M$ contains the information of the interband effects due to the crystalline SOI. The formula of the ratio $M$ given there is so general that it is applicable for various systems where the SOI plays an important role. Therefore, we can henceforth obtain new and rich informations of the crystalline SOI by measuring the ratio $M$ in various systems.

\begin{figure}
\begin{center}
\includegraphics[width=8cm]{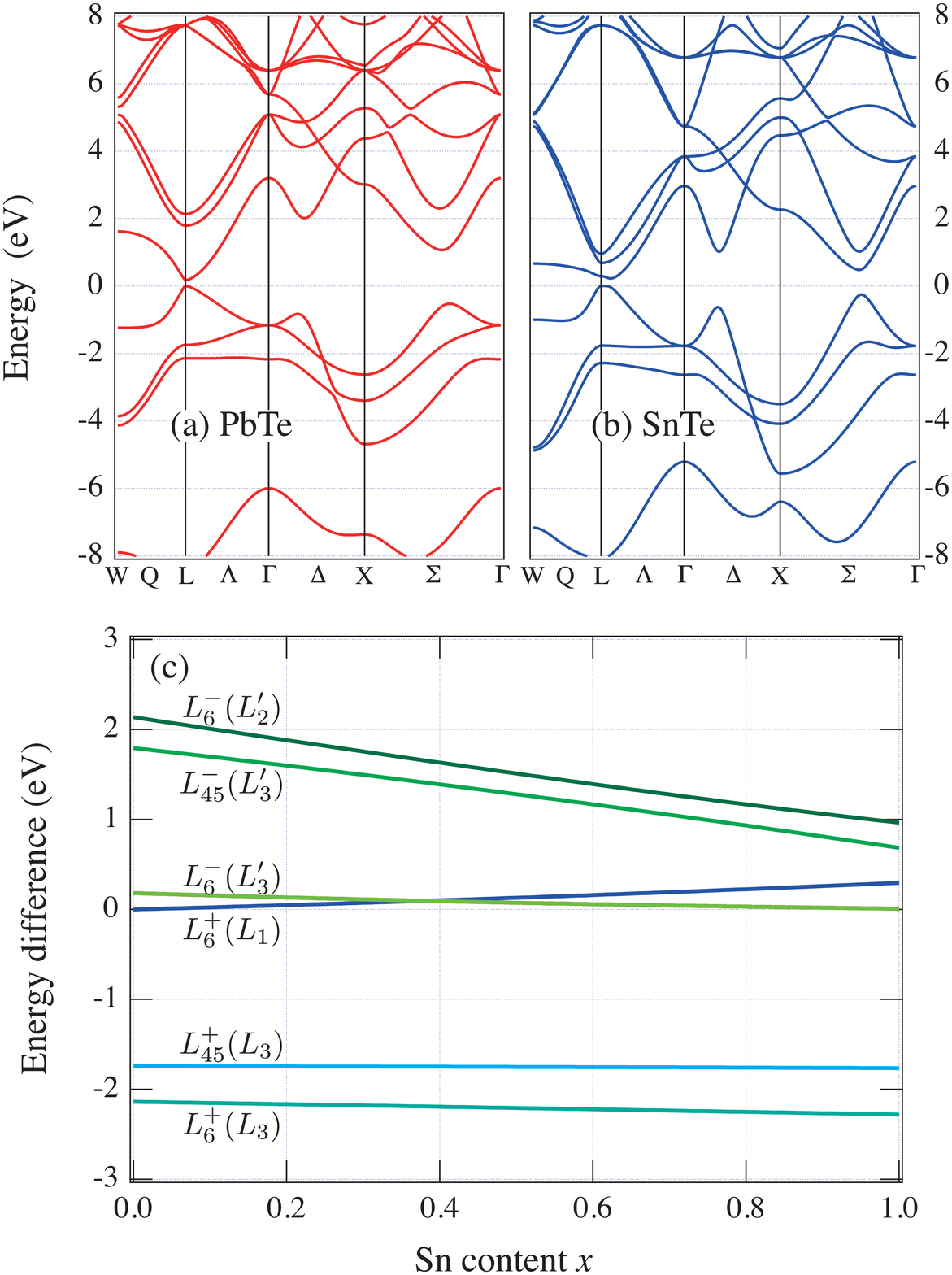}%
\caption{\label{Fig1}Band structures of (a) PbTe and (b) SnTe obtained by the relativistic empirical tight-binding calculation\cite{Lent1986}. (c)Positions of the energy band at the $L$-point as a function of Sn content ($x$) in Pb$_{1-x}$Sn$_x$Te calculated by the virtual crystal approximation.}
\end{center}
\end{figure}

	As another strongly spin-orbit coupled systems, the family of narrow-gap IV-VI semiconductors with rock-salt structure, such as PbTe, SnTe, GeTe, PbSe and SnSe, is of prime importance well recognized in the field of thermoelectronics\cite{Behnia_book} but newly interested in spin-orbit physics. Their alloy Pb$_{1-x}$Sn$_x$Te is the first known band inverted materials\cite{Dimmock1966,Dimmock1971,Lent1986}(Fig. \ref{Fig1}). The band inversion occurs since the ways of the band modification due to SOI are different between PbTe and SnTe. %Therefore, it is naturally expected that the modifications of the Zeeman splitting due to SOI are  also essentially different between the two. 
	Interestingly, the system turns from trivial to nontrivial topological crystalline insulators at this band inversion\cite{Fu2011,Hsieh2012}. Such a band inversion can affect the ratio $M$ through the interband effect of SOI, which has never  been recognized in other strongly spin-orbit coupled systems.
	
	In this paper, we report for the first time the intimate properties of the ratio $M$ in the band inverted system of Pb$_{1-x}$Sn$_x$Te based on the multiband $\kp$ theory with the relativistic empirical tight-binding band calculation. We show how the ratio $M$ changes by the band inversion. The obtained results establish the validity of the universal understandings obtained in Ref. \cite{Fuseya2015b}. As a by-product, it reveals that $M$ is useful not only to obtain the information of the crystalline SOI effect, but also to detect the band inversion point only by the bulk measurement of quantum oscillations. Furthermore, $M$ can be also used to evaluate quantitatively the ``Diracness" of the materials.

%\section{Theory}
	The multiband $\kp$ theory yields general formulas for the inverse mass tensor $\alpha$ and the g-factor $g$ in the forms\cite{Fuseya2015b}:
\begin{eqnarray}
	\alpha_{ij}&=\frac{\delta_{ij}}{m}+\sum_{n \neq 0}
	\frac{t_{n i}t_{n j}^* + u_{n i}u_{n j}^* + {\rm c. c.}}{E_0 - E_n},
	\label{eq1}
	\\
	g_i &=2m\sqrt{G_{ii}}, 
	\label{eq2}
	\\
	G_{ii}&= 4\left| \left( \sum_{n \neq 0}\frac{\bt_n \times \bu_n}{E_0 -E_n} \right)_i \right|^2
%	\nonumber\\&
	-\left( \sum_{n\neq 0} \frac{\bt_n \times \bt_n^* + \bu_n \times \bu_n^*}{E_0-E_n}\right)^2 ,
	\label{eq3}
\end{eqnarray}
	where $i$ and $j$ denote the directions, $E_0$ is the target band, and $E_n$ is the energy of the $n$-th nearest energy band from $E_0$. $\bt_n$ ($\bu_n$) is the interband matrix element of the velocity operator between the $0$-th and $n$-th band for the same (opposite) spins as $\bt_n = \langle \psi_{0 \uparrow}|\bm{v}|\psi_{n\uparrow} \rangle$ ($\bu_n = \langle \psi_{0 \uparrow}|\bm{v}|\psi_{n\downarrow} \rangle$). Equations \eqref{eq1}-\eqref{eq3} are general formulas to be valid for various materials. For the rocksalt structure of Pb$_{1-x}$Sn$_x$Te, we take the longitudinal axis ($z$) parallel to $(111)$-direction and the transverse axis ($x$ and $y$) perpendicular to it. 
	The ratios are then given by  
\begin{eqnarray}
	M_{\parallel}=\sqrt{\frac{G_{zz}}{\alpha_{xx}^2}}, \quad
	M_{\perp}=\sqrt{\frac{G_{xx}}{\alpha_{yy}\alpha_{zz}}},
	\end{eqnarray}
	where $M_\parallel$ and $M_\perp$ are for the magnetic field along the longitudinal and transverse directions, respectively.

	For two-band Hamiltonian only with $E_0$ and $E_1$, which is equivalent to the Dirac Hamiltonian, Eqs. \eqref{eq1}-\eqref{eq3} yields $M_\parallel = M_\perp = 1$, the common result of the two-band model\cite{Cohen1960,Wolff1964,Fuseya2015}. If we further take into account the contributions form the other band (i.e, more than three-band model), it was shown that the interband contributions from the higher (lower) energy bands decrease (increase) the ratio $M$ from unity for valence band\cite{Fuseya2015b}. Based on this, we can immediately speculate the behaviors of $M$ for Pb$_{1-x}$Sn$_x$Te without any specific calculations; this is a merit of the multiband $\kp$ theory. For PbTe, the top valence band of even parity ($L_6^+$) has finite matrix elements only between odd parity bands ($L_6^-$ and $L_{45}^-$), all of which locate above the valence band, so that $M$ should be less than unity. (The symmetry of each band at the $L$ point are shown in Fig. \ref{Fig1} (c).) For SnTe, on the other hand, the top valence band has odd parity ($L_6^-$), which couples with even parity bands ($L_6^+$, $L_{45}^+$). Then, the contribution from the lower energy bands become relevant resulting in $M$ larger than unity. It is naively expected that $M$ keeps unity for finite region in $x$, where the two band approximation is valid, but no one has examined both theoretically and experimentally.

%\section{Results}
%\subsection{Band calculation}
	First, we see the properties of $M$ calculated by the numerical band calculations. Here we adopt the relativistic empirical tight-binding model by Lent et al. \cite{Lent1986}. $s$-, $p$-, and $d$-orbitals are taken as the basis, i.e., 18 band model. (The number of the eigenenergy is 36 including the spin.) The band structures for PbTe and SnTe so obtained are shown in Fig. \ref{Fig1} (a) and (b).	
	In order to obtain the band structure of Pb$_{1-x}$Sn$_x$Te alloy, we employ the virtual crystal approximation, which demonstrates correctly the band inversion between conduction and valence bands\cite{Lent1986}. The energy levels at the $L$ point are shown in Fig. \ref{Fig1} (c) as a function of the alloy composition $x$. The band inversion occurs at $x=0.381$, which quantitatively agrees with the experimental observation\cite{Dimmock1966}. The matrix elements $\bt_n$ and $\bu_n$ are calculated for six bands [$L_{45}^+ (L_3), L_{45}^- (L_3'), L_6^+ (L_1), L_6^+ (L_3), L_6^- (L_3'), L_6^- (L_2')$] from the tight binding Hamiltonian. Then, $M$ is calculated just by substituting $\bt_n$, $\bu_n$, and $E_n$ into Eqs. \eqref{eq1}-\eqref{eq3}. Hereafter we focus on the properties of the top valence band assuming the p-type carries. Almost the same properties was obtained for the bottom conduction band. 

\begin{figure}[tb]
\begin{center}
\includegraphics[width=8cm]{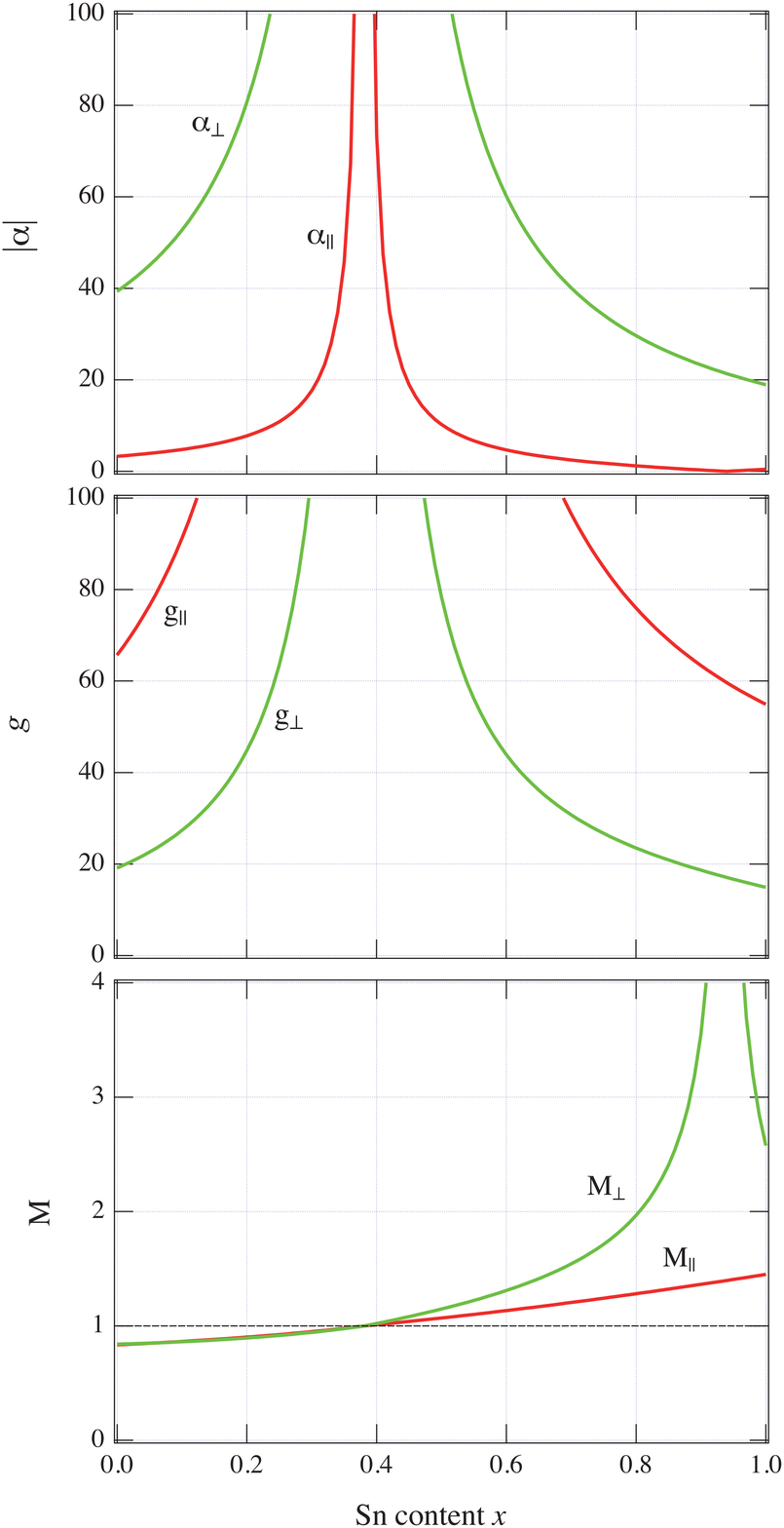}
\caption{\label{Fig3}Sn content dependences of (a) the inverse mass, (b) the g-factor, and (c) the ratio of the Zeeman splitting to the cyclotron energy for the longitudinal ($\parallel$) and the perpendicular ($\perp$) direction of the magnetic field.}
\end{center}
\end{figure}

The resultant $\alpha$, $g$, and $M$ are shown as a function of the Sn content $x$ in Fig. \ref{Fig3}. The results: $\alpha_\parallel= -3.40$, $\alpha_\perp=-40.4$, $g_\parallel=65.7$, and $g_\perp=19.1$ are in good agreement with the experimental values for PbTe\cite{Cuff1964,Bernick1970}.
There are four significant features of $M$: (i) $M_{\parallel, \perp}<1$ for PbTe and $M_{\parallel, \perp}>1$ for SnTe; (ii) $M_{\parallel,\perp}=1$ \emph{just} at the band inversion point; (iii) $M$ is very isotropic for PbTe side while it is anisotropic for SnTe side, and (iv) $M_\perp$ diverges in the SnTe side. The property (i) is consistent with the anticipation above, whereas properties (ii)-(iv) are the results that are firstly revealed by the present calculation. These properties can be clearly interpreted based on the group theoretical analysis in the following.

%\subsection{Group theoretical analysis}
	%First, we take into account the group theoretical symmetries at the $L$ point and study the qualitative properties of $M$. 
%	The symmetry of each band at the $L$ point are shown in Fig. \ref{Fig1}. The valence band of PbTe has the symmetry $L_6^+$, whereas that of SnTe has the symmetry $L_6^-$. Hereafter we concentrate on the properties of the valance band assuming the p-type carriers.
	
	For PbTe, the matrix elements with the $L_6^+$ top valence band are given as\cite{Dimmock1964}
	\begin{eqnarray*}
	\bm{t}_{61}&=\bra{L^{+}_{6}(L_1)}\bm{v}\ket{L^{-}_{6}(L_3')}=(a, ia, 0)
	,
	\\
	\bm{u}_{61}&=\bra{L^{+}_{6}(L_1)}\bm{v}\ket{C L^{-}_{6}(L_3')}=(0, 0, b) 
	,
	\\
	\bm{t}_{45}&=\bra{L^{+}_{6}(L_1)}\bm{v}\ket{L^{-}_{45}(L_3')}=(c,-ic,0) 
	=\bm{u}_{45},
%	$
%	\bm{u}_{45}=\bra{L^{+}_{6}(1)}\bm{\pi}\ket{C L^{-}_{45}(3')}=(c,ic,0) 
%	$,
	\\
	\bm{t}_{62}&=\bra{L^{+}_{6}(L_1)}\bm{v}\ket{L^{-}_{6}(L_2')}=(0, 0, d) 
	,
	\\
	\bm{u}_{62}&=\bra{L^{+}_{6}(L_1)}\bm{v}\ket{C L^{-}_{6}(L_2')}=(e, ie, 0)
	,
	\end{eqnarray*}
where $a\sim e$ are complex numbers, $C$ is the product of space inversion and time-reversal operatores. $L_{1,2,3}^{(')}$ denote the band symmetries without the SOI. %(The group theoretical notations are that of Ref. \cite{Dimmock1964}.) 
A straightforward calculation yields
\begin{eqnarray}
	\alpha_{xx}&=\alpha_{yy}=2\left(\frac{|a|^2}{\Delta_{1}}+\frac{2|c|^2}{\Delta_{2}}+\frac{|e|^2}{\Delta_{3}}\right),
	\label{eq5}
	\\
	\alpha_{zz}&=2\left(\frac{|b|^2}{\Delta_{1}}+\frac{|d|^2}{\Delta_{3}}\right),
	\label{eq6}
	\\
	G_{xx} &=G_{yy}=4\left(\frac{|ab|^2}{\Delta_{1}^2}+\frac{|de|^2}{\Delta_{3}^2}-\frac{abd^*e^*+a^*b^*de}{\Delta_{1}\Delta_{3}}\right), 
	\label{eq7}
	\\
	G_{zz}&=4\left(\frac{|a|^2}{\Delta_{1}}-\frac{2|c|^2}{\Delta_{2}}+\frac{|e|^2}{\Delta_{3}}\right)^2 ,
	\label{eq8}
\end{eqnarray}
where $\Delta_n = E_0 - E_n$ and $\Delta_3 < \Delta_2 < \Delta_1 <0$ for PbTe. Here, although we take into account whole six bands originating from the $p$-orbitals, only the contributions from the four bands ($E_{0\sim3}$) are finite. We obtain $M_{\parallel}=M_{\perp}=1$ if we take into account the contributions only from $E_0$ and $E_1$.

%{\it (i) $M$ less or greater than unity.}
(i)
The contributions from the higher energy bands, $E_{n\ge 2}$, lowers both $M_{\parallel}$ and $M_{\perp}$. For example, with the three-band model (taking into account $E_{0}$, $E_1$, and $E_2$ bands), we obtain very simple but exact formulas as
\begin{eqnarray}
	M_\parallel &= \frac{1-\lambda |X|^2}{1+\lambda|X|^2}, 
	\label{anisotropy1}\\
	M_\perp &=\frac{1}{\sqrt{1+\lambda |X|^2 }},
	\label{anisotropy2}
\end{eqnarray}
where $X=\sqrt{2}c/a$ and $\lambda = \Delta_1 / \Delta_2$ is a positive small value for PbTe. These results show that both $M_{\parallel}$ and $M_\perp$ decrease from unity due to the third band ($E_2$) contribution. 
For SnTe, the analytic forms for $\alpha_{ij}$, $G_{ii}$, and $M$ are the same as Eqs. \eqref{eq5}-\eqref{eq8} except for the different values of $a \sim e$. In the SnTe case, however, $\Delta_1 < 0 < \Delta_2 < \Delta_3$ due to the band inversion, so that $\lambda$ becomes negative resulting in the increase of $M$. These results verify the general understandings of $M$\cite{Fuseya2015b} in a clearer manner. 

%{\it (ii) $M=1$ just at the inversion point}
(ii)
Furthermore, Eqs. \eqref{anisotropy1} and \eqref{anisotropy2} give an exact proof of another important property of $M$ --- both $M_\parallel$ and $M_\perp$ cross unity \emph{just} at the band inversion point ($\lambda=0$). This is rather surprising since it is naively believed so far that $M=1$ (or the nontrivial Berry's phase\cite{Mikitik1999,Novoselov2005,Murakawa2013}) is something ``quantized" value and so it should hold for a finite range where the band gap is small and the two-band approximation is valid. However, Eqs. \eqref{anisotropy1} and \eqref{anisotropy2} prove that $M$ is a continuous value and not quantized for the actual materials, where there are always small but finite contributions from the other bands. $M=1$ is true only when the band gap $\Delta_1$ is zero.

\begin{figure}[t]
\begin{center}
\includegraphics[width=8cm]{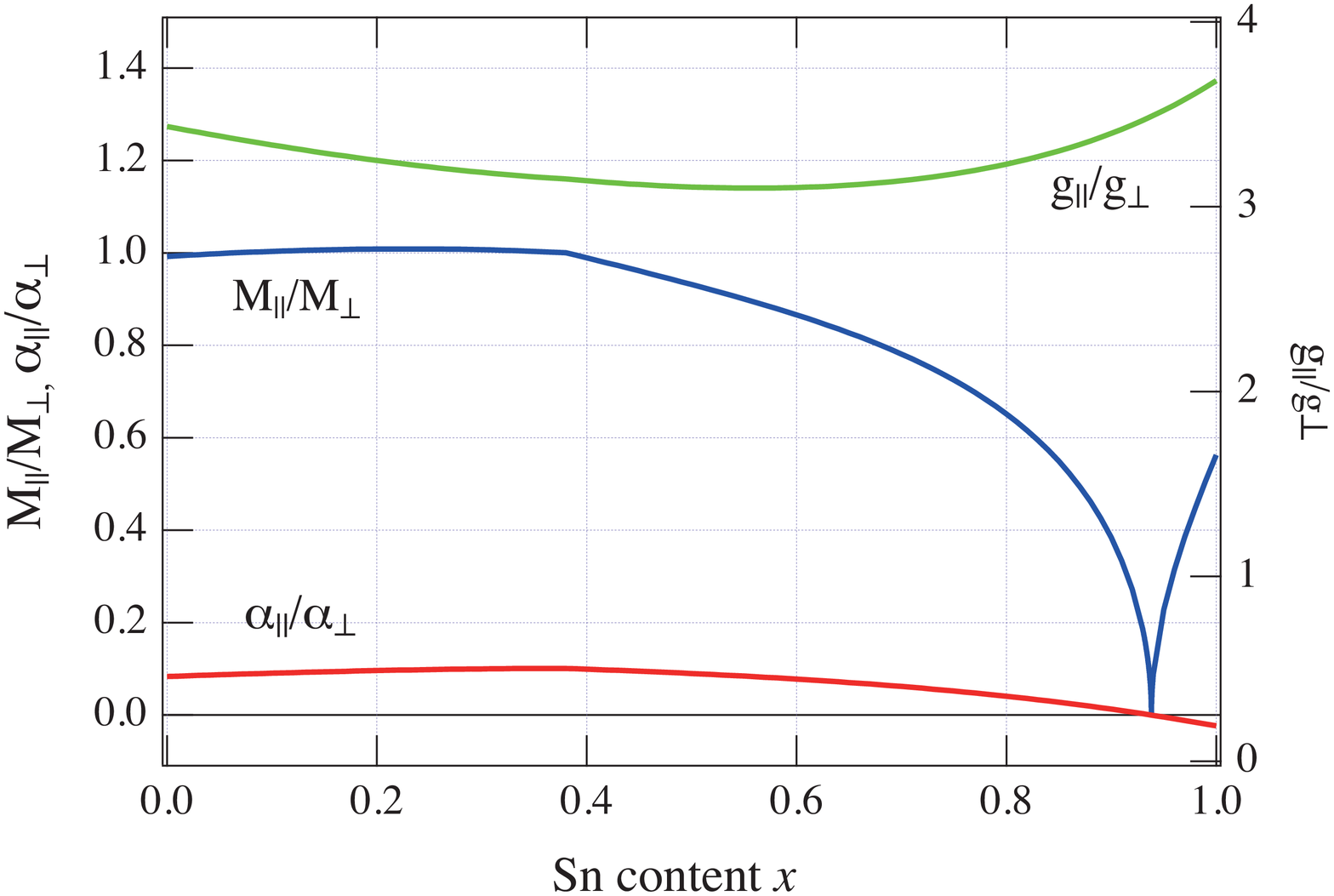}
\caption{\label{Fig4}Anisotropy of the inverse mass tensor $\alpha_{\parallel}/\alpha_{\perp}$, the g-factor $g_{\parallel}/g_\perp$, and the Zeeman-cyclotron ratio $M_{\parallel}/M_\perp$.}
\end{center}
\end{figure}

(iii) Figure \ref{Fig4} shows the Sn content dependence of $\alpha_\parallel / \alpha_\perp$, $g_\parallel / g_\perp$, and $M_\parallel / M_\perp$, which represents their anisotropy. It is clearly seen from Fig. \ref{Fig4} that $M$ in the PbTe side is very isotropic, while it is highly anisotropic in the SnTe side. From the three-band analysis of Eqs. \eqref{anisotropy1} and \eqref{anisotropy2}, the anisotropy of $M$ is 
\begin{eqnarray}
\frac{M_\parallel}{M_\perp} = \frac{1-\lambda|X|^2}{\sqrt{1+\lambda|X|^2}}\simeq 1-\frac{3}{2}\lambda|X|^2.
\end{eqnarray}
The anisotropy $M_\parallel / M_\perp$ should deviate from unity away from the band inversion point, which is incompatible with the numerical result. If we further take into account the whole six bands, we obtain in the exact forms as
\begin{eqnarray}
	M_\parallel &=& \frac{1-\lambda |X|^2 + \lambda' |Y|^2}{1+\lambda |X|^2 + \lambda' |Y|^2},
	\label{eq11}
	\\
	M_\perp &=&\frac{|1-\lambda' YZ^*|}{\sqrt{(1+\lambda'|Z|^2)(1+\lambda|X|^2+\lambda'|Y|^2)}},
	\label{eq12}
\end{eqnarray}
where $Y=e/a$, $Z=d^*/b^*$, and $\lambda'=\Delta_1 / \Delta_3$. For small $\lambda$- and $\lambda'$-terms, the anisotropy of $M$ can be written up to the six-band model as 
\begin{eqnarray}
	\frac{M_{\parallel}}{M_\perp}\simeq 1-\frac{3}{2}\lambda|X|^2 + \frac{1}{2}\lambda' |Y+Z|^2.
	\label{eq13}
\end{eqnarray}
%\begin{eqnarray}
%	M_\parallel =1,
%\end{eqnarray}
%for both PbTe and SnTe side, while
%\begin{eqnarray}
%	M_\perp \le 1 \quad\mbox{(for PbTe side)}\\
%	M_\perp \ge 1 \quad\mbox{(for SnTe side)}
%\end{eqnarray}
Hence, the contributions from $E_3$ can compensate with that from $E_2$, so that $M$ of PbTe should be considerably isotropic, though the surprisingly isotropic $M$ of the present result will be due to an accidentally perfect compensation. Actually, if we calculate $M$ from other tight-binding model by Lach-hab {\it et al.}\cite{Lachhab2000}, we obtained $M_\parallel/M_\perp = 0.909$; still isotropic, but not so perfectly. (Note that the expansion of Eq. \eqref{eq13} is valid for PbTe side, but invalid for SnTe side, where $M_\perp$ diverges. ) %Also note that, even in our calculation, $M_\parallel /M_\perp = 1.0***$ at $x=0$ with the $\delta_{ij}/m$-term in \eqref{eq1}, whereas $M_\parallel /M_\perp = 1.1***$ without it.

(iv)
Equations \eqref{eq11} and \eqref{eq12} also prove that $M_{\parallel, \perp}$ can diverge at some point of $\lambda, \lambda' <0$, i.e., in the SnTe side and never diverge in the PbTe side. There are two possibilities: (1) both $M_\parallel$ and $M_\perp$ diverge due to $1+\lambda |X|^2 + \lambda' |Y|^2 \to 0$; (2) only $M_\perp$ diverges due to $1+\lambda' |Z|^2 \to 0$. In the present calculation, only $M_\perp$ diverges at $x=0.938$, which means $1+\lambda' |Z|^2$ diverges (i.e., $\alpha_{zz}\to 0$). By analyzing the properties of both $M_\parallel$ and $M_\perp$, we can obtain the detailed informations of the interband matrix elements due to the crystalline SOI.

One of the most useful and practical aspects of the ratio $M$ is to be able to compare different materials through the common index $M$. It is worth noting that, for the holes at $T$ point in Bi, $M \sim 2$ for one direction of the magnetic field, and $M=0$ for the perpendicular direction\cite{Smith1964,Edelman1976,Bompadre2001,Behnia2007_PRL,ZZhu2011,Fuseya2015b}. These properties are very different from the isotropic $M\sim 1$ in PbTe, although the $T$ point of rhombohedral Bi has the equivalent symmetries to the $L$ point of rocksalt PbTe\cite{Dimmock1964}. This difference can be interpreted as the difference of the order of the band: the hole band of Bi has the $T_{45}^-$ symmetry, which is equivalent to $L_{45}^-$, whereas that of PbTe has the $L_{6}^+$. Only this difference causes the significant difference in the property of $M$. In other words, $M$ is very sensitive to the symmetry of the band and the interband matrix elements that is crucial for the crystalline SOI.

The initial purpose of the present work was to study the ratio $M$ for the specific model of Pb$_{1-x}$Sn$_x$Te alloys and enrich the understanding of the general properties of $M$. The present results, however, give rise to byproducts.
First, by using the property (ii), we can detect the exact point of the band inversion through the ``bulk" measurement of $M$, such as the quantum oscillation measurements. This means the bulk detection of the transition from the trivial to the non-trivial topological crystalline insulator\cite{Fu2011,Hsieh2012}. This approach will be a complement to the optical spectroscopy measurements.

Second, we can quantitatively evaluate the ``Diracness" in terms of $M$. Nowadays various type of materials have been recognized as the Dirac electron systems, whose effective Hamiltonian is given by the Dirac Hamiltonian, e.g., Bi\cite{Wolff1964,Fuseya2015}, PbTe, Ca$_3$PbO\cite{Kariyado2011} etc. Naively, when the band dispersion of the candidate material looks like Dirac dispersion, one may call it Dirac electron systems. But the judgment of this kind is too sensual. If one uses $M$ as the indicator of the Dirac electrons, one can make a quantitative comparison of the Diracness between different materials. For example, $M=1.00$-$1.02$ for electron carrier in Bi with the field along the bisectrix axis\cite{ZZhu2011}, while $M = 0.834$ for PbTe.

In summary, we studied the ratio of the Zeeman splitting to the cyclotron energy, $M=\Delta E_Z /\hbar \wc$, for Pb$_{1-x}$Sn$_x$Te based on the multiband $\kp$ theory. By employing the tight-binding band calculation, We calculated $\alpha$, $g$, and $M$ numerically. Based on the group theoretical analysis, we obtained the simple but exact formulas of $M$, \eqref{eq11} and \eqref{eq12}, which deepen the understandings of $M$. We certified that $M<1$ for PbTe and $M>1$ for SnTe. It is rather surprising that the band 2 eV far from the Fermi energy gives a sizable contribution to $M$; a strong indication of the large interband coupling due to the crystalline SOI. We found that $M$ crosses unity just at the band inversion point. By using this property, we can detect the transition to the topological crystalline insulator only from the bulk measurement of quantum oscillation. We can obtain detailed informations of the interband matrix elements due to the crystalline SOI by analyzing both $M_\parallel$ and $M_\perp$. Also, $M$ can be used as a good indicator that can quantitatively evaluate how the electrons in crystal is close to the Dirac electrons.

\ack
We thank K. Behnia and B. Fauqu\'e for fruitful discussions.
This work is supported by JSPS KAKENHI Grants No. 25870231.
\\

\bibliographystyle{iopart-num}
\bibliography{PbTe}

\end{document}